\documentclass[twocolumn,superscriptaddress,groupedaddress]{revtex4}  

\usepackage{graphicx}  
\usepackage{dcolumn}   
\usepackage{bm}        
\usepackage{amssymb}   

\begin{document}


\title{Brillouin amplification supports $1\times10^{-20}$ accuracy in optical frequency transfer over 1400~km of underground fibre}

\author{Sebastian M. F. Raupach}\email{sebastian.raupach@ptb.de}
\author{Andreas Koczwara}
\author{Gesine Grosche}
\affiliation{Physikalisch-Technische Bundesanstalt (PTB), Bundesallee 100, D-38116 Braunschweig, Germany}

\begin{abstract}
We investigate optical frequency transfer over a 1400~km loop of underground fibre connecting Braunschweig and Strasbourg. Largely autonomous fibre Brillouin amplifiers (FBA) are the only means of intermediate amplification, allowing phase-continuous measurements over periods up to several days. Over a measurement period of about three weeks we find a weighted mean of the transferred frequency's fractional offset of $(1.1\pm0.4)\times10^{-20}$. In the best case we find an instability of $6.9\times10^{-21}$ and a fractional frequency offset of $4.4\times10^{-21}$ at an averaging time of around 30~000~s. These results represent an upper limit for the achievable uncertainty over 1400 km when using a chain of remote Brillouin amplifiers, and allow us to investigate systematic effects at the $10^{-20}$-level.
\end{abstract}

\date{\today}

\maketitle


The transfer of optical frequencies \cite{lopez2012frequency,predehl2012,lopez2013,droste2013,calonico2014,raupach2014} is required e.g. for comparing the ultrastable frequency signals generated by optical atomic clocks. Latest reports state record clock instabilities and accuracies on the $10^{-18}-$level \cite{bloom2014,ushijima2015,nicholson2014arxiv}. Transferring their ultrastable optical frequency, as required for clock validation via remote comparison as well as for applications in relativistic geodesy \cite{vermeer1983,chou2010,delva2013}, is challenging. Optical frequency transfer can also serve to disseminate an optical reference frequency to remote users \cite{raupach2014,pape2010}. Taking advantage of phase synchronicity at the sending and receiving end within the bandwidth of the link stabilization \cite{raupach2014b}, it can be used for the common-mode suppression of the Dick-effect in optical clock comparisons via synchronous sampling \cite{ushijima2015,takamoto2011,akatsuka2014}.\\
The noise added by the transmission path can be minimized by using underground fibre links and be suppressed further by actively stabilizing the phase of the transferred signal \cite{predehl2012,lopez2012frequency,calonico2014}. Alternatively the two-way technique can be employed \cite{calosso2014twoway,bercy2014,stefani2014arxive}, where applicable. \\
The fibre's attenuation is traditionally compensated for by broad-band Erbium doped fibre amplifiers (EDFA). For ultrastable optical frequency transfer they have to be operated without isolators to ensure symmetry of the optical path in both directions \cite{sliwczynski2013,raupach2013eftf}. This bidirectional operation typically restricts the gain to around 17 dB to avoid spontaneous lasing.\\
Over long distances, this leads to excess losses, and in some long-distance, bidirectional EDFA links the signal phase is lost at least several times per hour \cite{droste2013,calonico2014}. Very recently, an instability (as estimated by the modified Allan deviation) of $3\times10^{-20}$ was reported for phase-stabilized frequency transfer over a 540 km loop of cascaded fibre links \cite{stefani2014arxive}; an accuracy was not reported.\\
However, in view of clock (in-)accuracies on the $10^{-18}$-level, hitherto demonstrated inaccuracy contributions of long-distance optical frequency transfer of several times $10^{-19}$ \cite{predehl2012,droste2013,lopez2013,calonico2014,raupach2014} can no longer be neglected.\\
Here we present results from a measurement campaign performed over a 1400 km fibre loop from Braunschweig (Germany) to Strasbourg (France) and back. This fibre link will become part of the future international link PTB (Germany) / LNE-SYRTE (France). Its length corresponds to twice the geographical distance Sydney-Brisbane and to the \emph{total} length of the link PTB / LNE-SYRTE.\\
In this Letter we report a signal amplification approach adapted to ultrastable-frequency metrology, which allows phase-continuous measurements over periods up to several days. We show that it supports an instability and inaccuracy better than $10^{-20}$. We introduce the $\Lambda$Allan deviation ($\Lambda$ADEV) as an improved measure of uncertainty, well suited for remote clock comparisons. Using separately housed setups at the local and ``remote'' end, we demonstrate the transfer of an optical frequency over 1400~km with an inaccuracy of around $2\times10^{-20}$.\\
\begin{figure*}
\includegraphics[width=13cm]{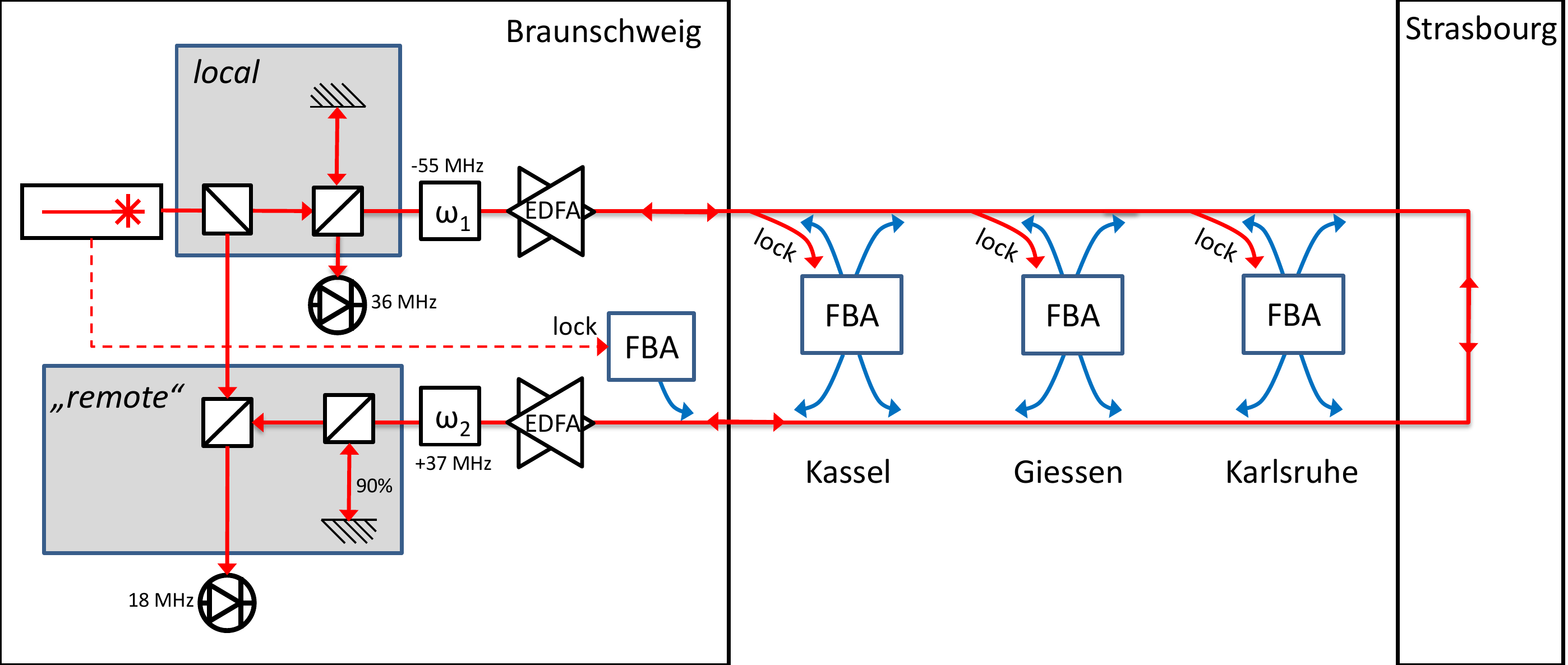}
\caption{Schematic sketch of the the measurement setup and the 1400~km Brillouin link PTB - Universit\'e de Strasbourg - PTB.}
\end{figure*}
The link consists of a pair of telecommunication fibres connecting PTB in Braunschweig to Universit\'e  de Strasbourg (UDS)over a fibre length of around 710~km. The fibres are patched together at UDS to form a loop of around 1400 km length, where a cascade of fieldable fibre Brillouin amplifiers (FBA) is the only means of intermediate amplification. These have been developed at PTB after early demonstrations of Brillouin-amplification assisted optical frequency transfer \cite{terra2010,predehl2012}. In contrast to EDFA, the FBA also in bidirectional operation allow a gain of 40 dB and more, without losing the signal phase over periods up to several days \cite{raupach2013eftf,raupach2014}.\\The fibre loop starts and ends in the same laboratory at PTB to facilitate the characterisation of the link. For active phase stabilization, the link is set up as a fibre Michelson interferometer. As a frequency source, we use a laser at 194.4 THz locked to a cavity-stabilized 1~Hz master laser. The local interferometer used for stabilization of the link, and the setup for detecting the ``remote''  beat frequency between the transferred and the source light are located in separate but adjacent, thermally insulated housings.  We use bidirectional EDFA as booster amplifiers for the outgoing and the incoming light. Beat frequencies are recorded using K\&K FXE dead-time-free totalizing counters with an internal gate time of 1 ms \cite{kramer2004}. The counters are set to report the frequency values at 1 second intervals. The frequency values are reported both as unweighted averages ($\Pi$-mode) and as triangularly weighted, overlapping averages (phase-averaging- or ``$\Lambda$''-mode \cite{dawkins2007}). Due to the spectral response of the triangle function, the $\Lambda$-type averaging more strongly suppresses Fourier frequencies larger than the reciprocal of the gate time, acting like a low pass filter (see \cite{dawkins2007,williams2008} for a related discussion). The overall measurement setup at PTB is described in \cite{raupach2014}.\\
Along the link three fieldable FBA are the only means of intermediate amplification. They are located in server rooms of Kassel University, Giessen University and the Karlsruhe Institute of Technology. While they do allow remote control, they normally operate autonomously. During one pass through this loop the signal experiences seven Brillouin amplifications, corresponding to an average inter-amplifier distance of 200~km. The individual lengths bridged and their attenuations are 205~km/44~dB (Braunschweig-Kassel), 160~km/33~dB (Kassel-Giessen), 231~km/47~dB (Giessen-Karlsruhe) and around $2\times114$~km/54~dB (Karlsruhe-Strasbourg-Karlsruhe), respectively. Note that the first fibre stretch has been shortened slightly since the measurements described in \cite{raupach2014}.\\Each FBA contains one pump laser, the light of which is split into four paths, where in two of the paths the frequency is shifted by one acousto-optic modulator to accommodate the frequency shift of the signal returning from the remote end. Thus each FBA simultaneously amplifies the signal in both directions per fibre and in both fibres of the loop. The frequency of the pump laser is locked to the incoming, amplified signal at an offset frequency corresponding to the fibre's Brillouin frequency ($\approx11$~GHz).  The FBA allow remote optimization of the pump light’s polarization and the offset lock frequency. Without polarization adjustment, all FBA stay in lock at least several days, often about one week or more. Each FBA can be relocked remotely.\\
To assess the accuracy of the optical frequency transfer, we calculate the unweighted mean of the difference between the expected and the measured beat frequency at the ``remote'' end.\\
We have performed two measurement campaigns from Decembre 10, 2014 to Decembre 18, 2014, and from Decembre 23, 2014 to January 2, 2015, covering a period of about three weeks. During the second campaign the stabilization of the link failed during five 1-second-intervals, during the first campaign we found an average of about four invalid data points per day. This demonstrates a high reliability and very low cycle slip rate. During all campaigns, the FBA were running unattended.\\
\begin{figure}
\includegraphics[width=9cm]{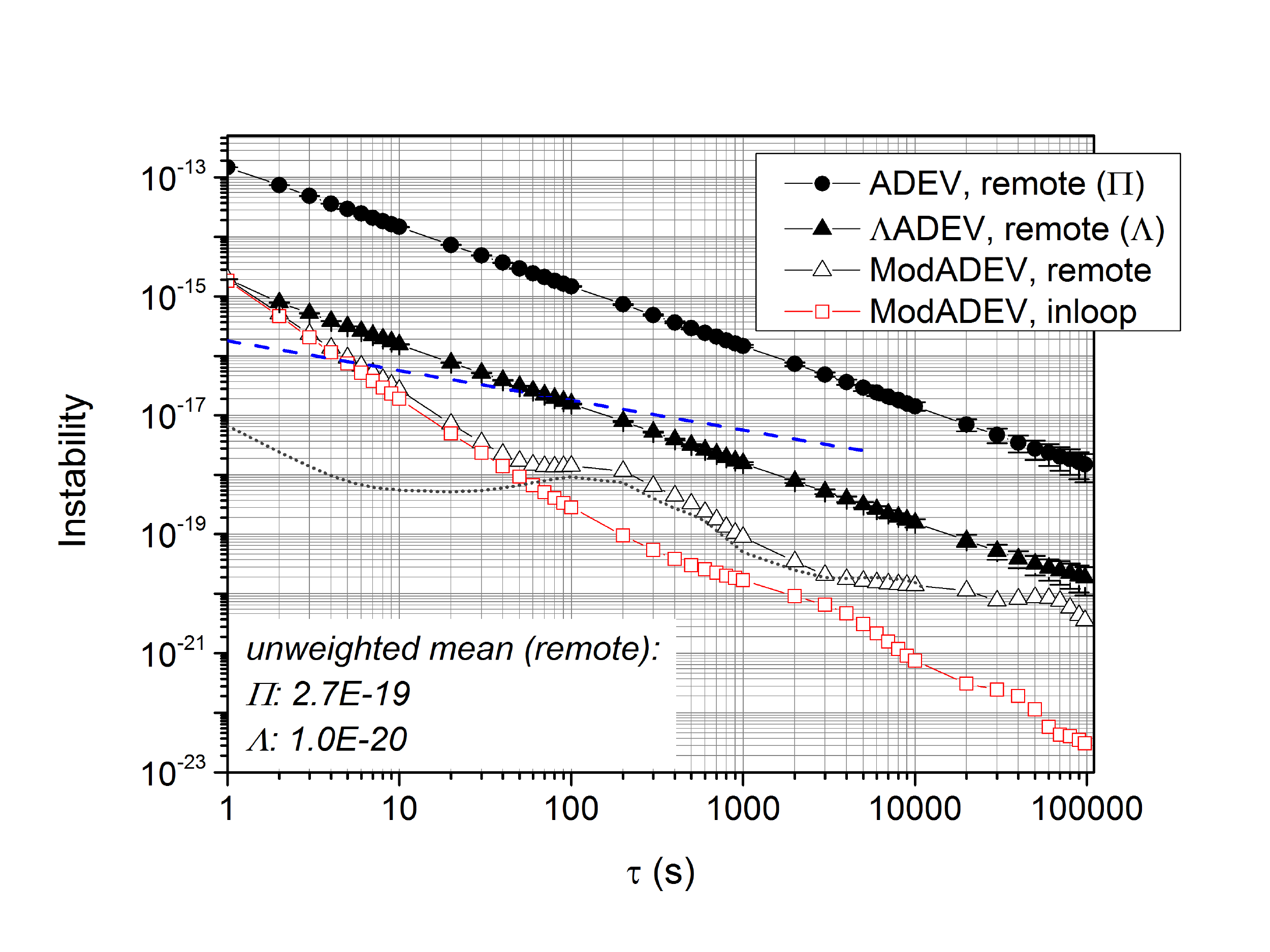}
\caption{Instability of the optical frequency transfer over the stabilized 1400 km ''Brillouin-Link'' Braunschweig-Strasbourg-Braunschweig. The frequency data are reported once per second simultaneously in $\Pi$- and in $\Lambda$-mode using totalizing dead-time-free counters (K\&K); filled circles: Allan deviation of the frequency transfer ($\Pi$-type data); filled triangles: Allan deviation formalism applied to $\Lambda$-type remote frequency offset data ($\Lambda$ADEV); open triangles modified Allan deviation of the transferred frequency; open squares: modified Allan deviation of the return signal (inloop signal); dashed line: instability of the most recent side-by-side optical clock comparisons \cite{ushijima2015,nicholson2014arxiv}; dotted line: instability without link (``optical short-cut''); the error bars of each ADEV point $\sigma_{\textrm{y}}(\tau)$ are calculated as $\sigma_{\textrm{y}}(\tau)/\sqrt{N(\tau)}$ \cite{riley2008}, with $N(\tau)$ being the number of averaging intervals at $\tau$.}
\label{fig2}
\end{figure}
Results for optical frequency transfer over the 1400~km are shown in figure 2. The filled circles illustrate the instability (Allan deviation, ADEV \cite{riley2008}) of unweighted averages of $\Pi$-data \cite{dawkins2007}. We find an unweighted mean of the transfer-induced frequency offset of $2.7\times10^{-19}$, well within the instability of $1.5\times10^{-18}$ indicated by the last value of the Allan deviation.\\
We apply the Allan deviation formalism also to the data reported in $\Lambda$-mode. The slope of $1/\tau$ of the instability curve remains unchanged. The overall instability is smaller than the ADEV of the $\Pi$-data, benefiting from the low-pass characteristic of $\Lambda$-type averaging. To distinguish this instability curve from the genuine ADEV, we label it $\Lambda$ADEV. Here we find an instability ($\Lambda$ADEV) at an averaging time $\tau$ of around 97~000~s of $1.9\times10^{-20}$, and an unweighted mean of the $\Lambda$-data of $1.0\times10^{-20}$.\\
Also shown are the modified Allan deviations (ModADEV \cite{dawkins2007,riley2008}) of the remote and inloop frequencies, calculated from the $\Lambda$-data. The modified Allan deviation effectively continues the triangular, overlapping $\Lambda$-type averaging over the averaging interval $\tau$, benefiting from its low-pass behaviour \cite{dawkins2007}. Contrary to the ADEV it distinguishes white phase noise from flicker phase noise and reveals noise processes otherwise covered by white phase noise. Thus the ModADEV helps to gain an approximate idea of the achievable uncertainty. For averaging times $\tau$ up to about 10~s, the ModADEV drops as $1/\tau^2$ \cite{droste2013}.\\For averaging times beyond around 100~s the ModADEV is similar to the differential instability between the separately housed local and ``remote'' setup, which shows some variability in the $10^{-20}$ range. In fig. \ref{fig2} we show one measurement of this relative instability, obtained by optically short-cutting the link (ModADEV, dotted line).\\
The dashed line in fig. \ref{fig2} indicates the instability of the most recent side-by-side or self-comparison of optical clocks \cite{ushijima2015,nicholson2014arxiv}. It illustrates that using $\Pi$-type data for a clock comparison would yield an unweighted frequency average dominated by the instability of the transfer up to the longest averaging times. This is in contrast to $\Lambda$-type data, where for averaging times larger than about 100~s the optical clocks start to dominate the instability.
Depending on the application, we may increase the $\Lambda$-type averaging interval to exploit the strong decrease in instability, thus ``sliding down'' along the ModADEV curve for continuous measurements (see fig. \ref{fig3}). For comparing the mean frequencies of e.g. optical clocks, the $\Lambda$-type averaging interval may be maximized. This achieves a lower instability ($\Lambda$ADEV) and thus a better estimate of the unweighted average.\\
\begin{figure}
\includegraphics[width=9cm]{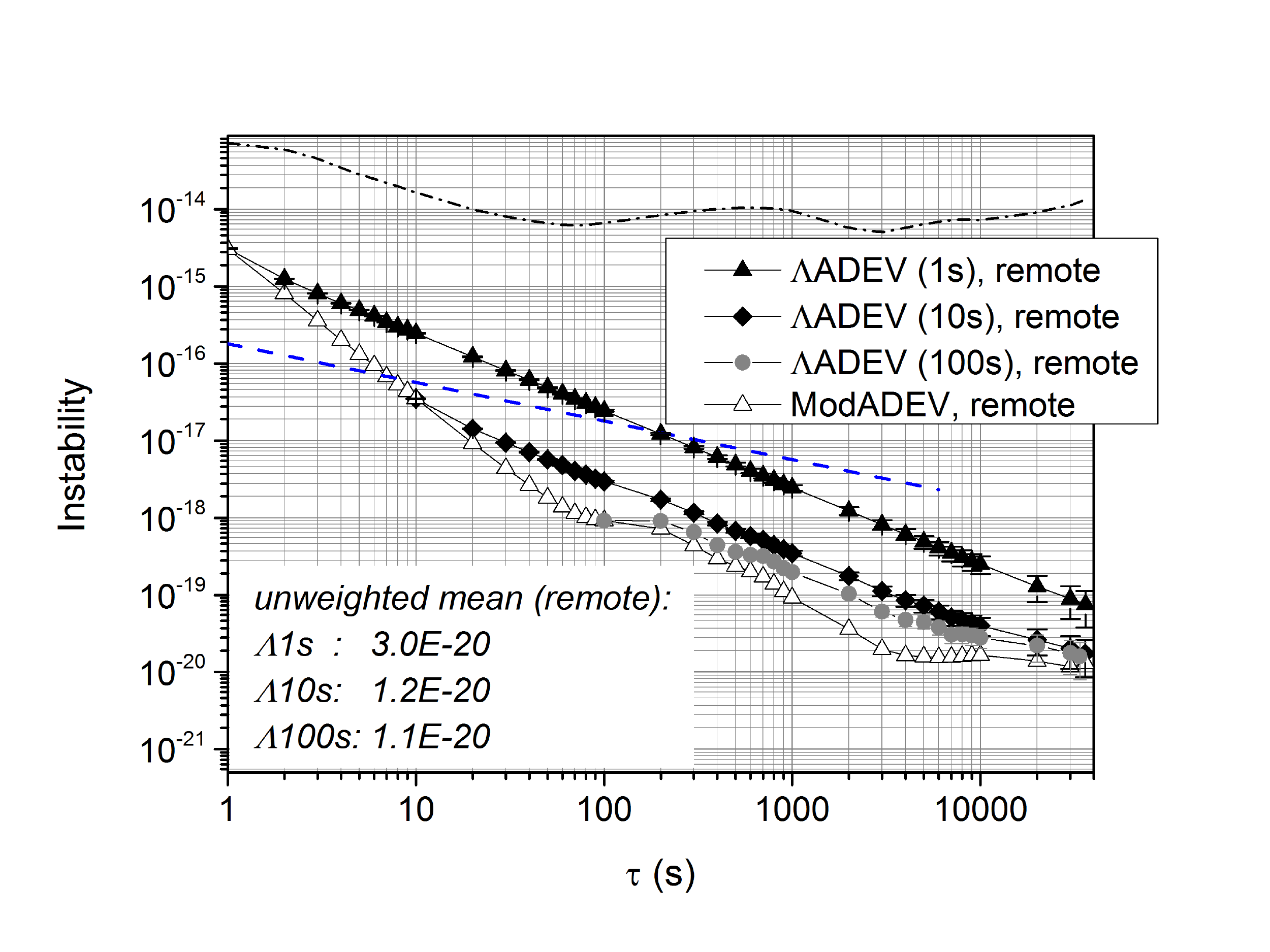}
\caption{Optical frequency transfer instability, different measurement and demonstration of a numerically extended $\Lambda$ averaging interval. Filled triangles: $\Lambda$ADEV of the frequency transfer for $\Lambda$-type data averaged over 1s; filled diamonds: $\Lambda$ADEV of the frequency transfer for $\Lambda$-type data, where the time for $\Lambda$-type averaging has been extended numerically to 10~s (grey circles: 100~s); open triangles: modified Allan deviation of the transferred frequency; dashed line: instability of optical clock comparison as in fig. \ref{fig2}; dash-dotted line: ModADEV of the fiber link stabilization's correction signal: the fiber noise is suppressed by six orders of magnitude for the longest averaging time.}
\label{fig3}
\end{figure}
The effect of a $\Lambda$-extension is demonstrated in fig. \ref{fig3}. It shows data taken over a period of 145.000 seconds (measurement 3 in fig. \ref{fig4}, which covers a slightly smaller period due to our $\Lambda$-extension algorithm). At the longest averaging times the $\Lambda$ADEV of the 1~s $\Lambda$-data is about six times larger then the instability floor as estimated by the ModADEV. We note, that for the longest averaging time of around 36~000~s we observe a noise suppression by about six orders of magnitude, close to the predicted delay limit for $f=36~000^{-1} \textrm{s}^{-1}$ according to \cite{williams2008}.\\
We numerically increase the $\Lambda$-type averaging to an interval of 10~s, the typical minimum interval at which an optical clock's instability starts to average down. We obtain an unweighted mean of these 10~s $\Lambda$-data of $1.2\times10^{-20}$ and an instability ($\Lambda$ADEV) of $1.7\times10^{-20}$ at $\tau\approx36~000$~s. At this extended $\Lambda$-interval, the transfer instability is smaller than that reported for a \emph{local} clock comparison \cite{ushijima2015} already for averaging times of 10~s onwards.\\
For the data shown in fig. \ref{fig2}, numerically increasing the $\Lambda$-averaging interval to 10~s yields an instability ($\Lambda$ADEV) at the longest averaging time of $9.9\times10^{-21}$, and an unweighted mean of $1.2\times10^{-20}$. \\
\begin{figure}
\includegraphics[width=9cm]{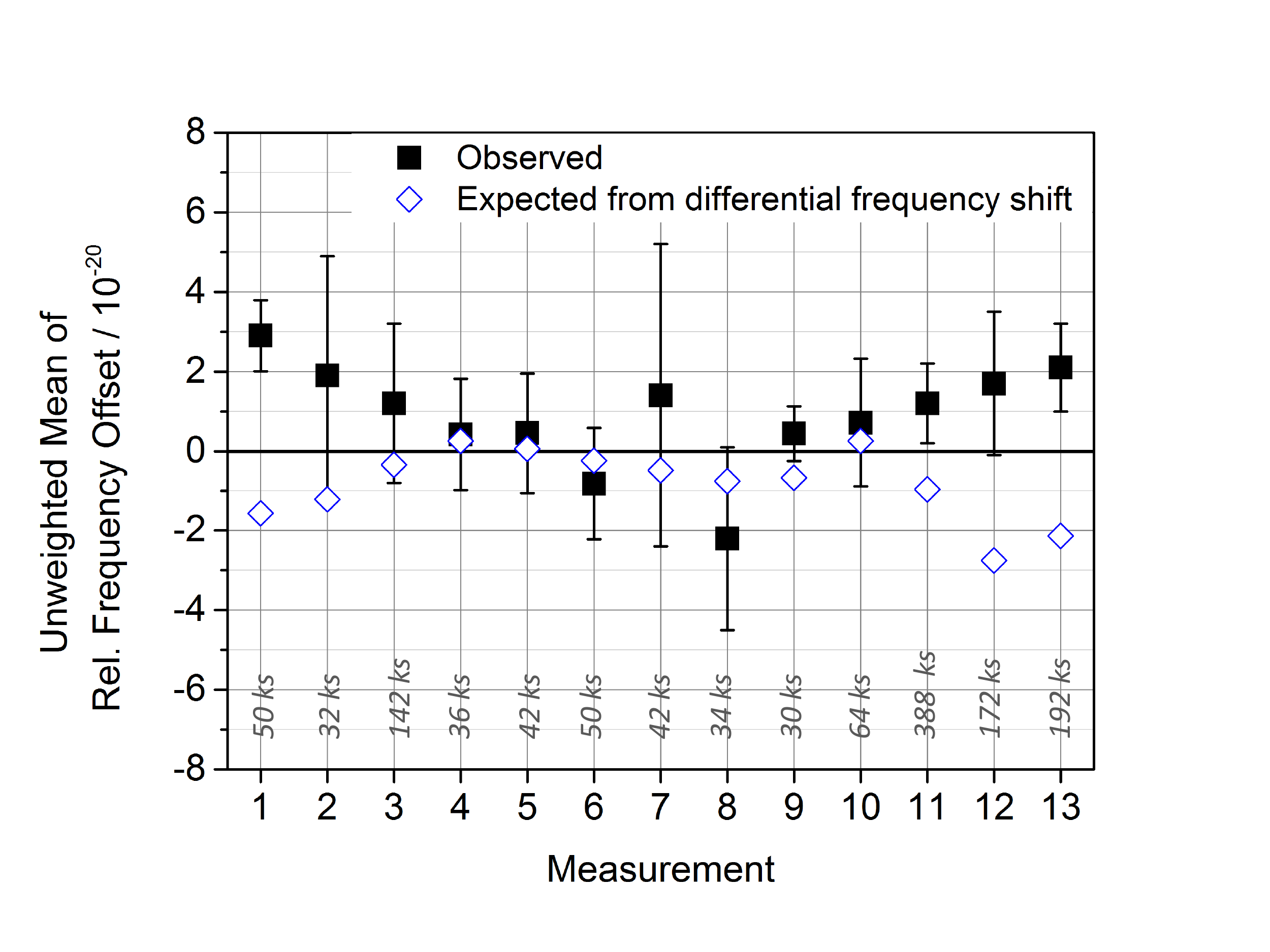}
\caption{Overview over all measurements during the period from Decembre 10, 2014 to January 2, 2015. The black squares indicate the unweighted means for numerically extended $\Lambda-$type averaging intervals of 2000~s, the respective measurement times are indicated in dark gray. The errors bars $\pm s$ indicate the last value $s$ of the corresponding $\Lambda$ADEVs. The mean of these data weighted by $1/s^2$ is $(1.1\pm0.4)\times10^{-20}$. For comparison, the open diamonds correspond to the unweighted mean of the fibre-induced frequency offset during the respective measurements, multiplied by a factor of $-2\times10^{-7}$. This factor corresponds to the relative frequency difference between outgoing and return light.}
\label{fig4}
\end{figure}
Fig. \ref{fig4} illustrates the results from all long-term measurements during the two measurement campaigns. Here, we numerically extended the $\Lambda$-type averaging interval to 2000~s for all measurements. The durations of the measurements are indicated in dark gray (in kiloseconds). The observed residual offsets scatter over a range of less than $\pm3\times10^{-20}$, their weighted mean is $(1.1\pm0.4)\times10^{-20}$. In the best case, during a measurement period of around 30~000~s, we obtain an instability ($\Lambda$ADEV) of $6.9\times10^{-21}$ and an unweighted mean of the transferred frequency's offset of $4.4\times10^{-21}$. We find, that the offset of the transferred frequency during some measurements significantly differs from Zero by up to $2\times10^{-20}$. We do not observe a significant offset of the inloop beat frequency. Thus, the observed offsets most likely indicate out-of-loop processes or non-reciprocal effects along the link.\\Due to the $2\times37$~MHz shift by the ``remote'' AOM, the outgoing``remote'' light and the returning roundtrip light exhibit a relative frequency difference of $3.8\times10^{-7}$. As the frequency shift is positive on the return path, we expect an over-compensation of the frequency offset observed at the remote end by around $2\times10^{-7}$ \cite{lopez2010} (see also eq. \ref{eq:Brillouin-Temp}) as shown in figure \ref{fig4}. We do not observe the expected behaviour, indicating the presence of other, competing processes.\\One process could be the variation of the Brillouin gain curve relative to the signal with temperature, in particular as the ``remote'' FBA amplifies one direction only. To estimate this effect, we use the frequency deviation of the AOM employed for link stabilization as a measure of the temperature variation $\textrm{d}T/\textrm{d}t$ in the fibre according to:
\begin{equation}
\Delta\nu_{\textrm{Link}}=-\frac{1}{2\pi}\frac{\textrm{d}\phi}{\textrm{d}t}\approx -\frac{L}{\lambda}k_n\frac{\textrm{d}T}{\textrm{d}t},
\label{eq:Brillouin-Temp}
\end{equation}

where we neglected the effect of the thermal expansion of the fibre ($\delta L/\delta T$). We use a thermo-optic coefficient of $k_n=\delta n/\delta T=1.1\times10^{-5}/\textrm{K}$ \cite{pinkert2015}, a fibre length $L$ of $1.4\times10^6$~m and assume a temperature dependence of the Brillouin frequency of $k_T^B=1.1$~MHz/K \cite{parker1997}. To investigate the phase sensitivity $k_{\phi}$ of the signal to the Brillouin frequency variation we deliberately modulate the ``remote'' FBA pump frequency over a range of about 6~MHz at a rate of several tens of kilohertz per second. We find $k_{\phi}\approx -4\times10^{-7}$~rad/Hz. This yields a scaling factor of:
\begin{equation}
\frac{\Delta\nu^{\textrm{Brillouin}}_{\textrm{Signal}}}{\Delta\nu_{\textrm{Link}}}\approx -\frac{k_T^Bk_{\phi}\lambda}{2\pi k_nL} = 8.5\times10^{-9}.
\label{eq:Brillouin-Offset}
\end{equation}

This effect is too small to compensate for the expected effect of asymmetric frequencies and hence does not limit us yet. We conservatively estimate the overall effect of the Brillouin amplification on the instability to be smaller than $5\times10^{-21}$. From the noise spectrum the ``remote'' FBA's lock we expect its instability contribution to be at least one order of magnitude smaller.\\Other effects that may play a role are the finite suppression of the fibre noise \cite{williams2008} as well as the relative fluctuations between the separate local and ``remote'' setups. These are subject to further investigations.\\

Using a loop consisting of a patched pair of telecommunication fibres connecting Braunschweig to Strasbourg, we have investigated optical frequency transfer over 1400 km with three FBA as the only means of remote amplification. We have introduced the $\Lambda$ADEV as a more stringent measure of instability. We have found that Brillouin amplification supports an inaccuracy and instability better than $1\times10^{-20}$, where the statistical precision is indicated by the instability \cite{hinkley2013}. Using separately housed local and ``remote'' setups, over a period of around three weeks  we observe an (in-)accuracy of the transferred optical frequency of about  $(1.1\pm0.4)\times10^{-20}$. We do not observe the effect expected from the relative frequency shift between the outgoing and returning light.\\To our knowledge, the results presented here are the first demonstration of frequency transfer at the $10^{-20}$ accuracy level over a continental-scale distance. This fibre link supports remote comparisons of the world's best optical clocks. We note that for applications in relativistic geodesy, the achieved level of accuracy and stability would correspond to a relativistic height resolution of around 100~$\mu$m.\\We thank H. Schnatz and F. Riehle at PTB for their long-standing support and are indebted to Paul-Eric Pottie and colleagues for enabling the cooperation with Universit\'e de Strasbourg. We thank P. Gris, B. Moya and colleagues from UDS, O. Bier from ARTE,  T. Vetter,  H. Klatte, and U. Koc from Kassel University, K. Ackermann and colleagues from Justus Liebig University in Giessen, B. Hoeft and colleagues from the Karlsruhe Institute of Technology, and F. Hack and colleagues from Gasline GmbH, as well as C. Grimm and colleagues from Deutsches Forschungsnetz e.V. (DFN) and Emilie Camisard from RENATER. We are grateful to Thomas Legero for operating the cavity-stabilized laser, and to Andre Uhde, J\"orn Falke, Mattias Misera, and Marion Wengel for excellent technical support.  This work is supported by the European Metrology Programme (EMRP) SIB-02 (“NEAT-FT”). The EMRP is jointly funded by the EMRP participating countries within EURAMET and the European Union.

\end{document}